\documentclass{article}

\pdfoutput=1
\usepackage[utf8]{inputenc} 
\usepackage[T1]{fontenc}    
\usepackage[hidelinks]{hyperref}
\usepackage[authoryear]{natbib}
\usepackage{url}            
\usepackage{booktabs}       
\usepackage{amsfonts}       
\usepackage{nicefrac}       
\usepackage{microtype}      
\usepackage{lipsum}
\usepackage{graphicx}
\usepackage{subcaption}
\usepackage{setspace}
\usepackage{multicol}
\usepackage{subcaption}
\usepackage{float}
\usepackage{tcolorbox}
\usepackage{colortbl}
\usepackage{multirow}
\usepackage{amsmath}
\usepackage{tabularx}
\makeatletter

\renewcommand{\normalsize}{%
  \@setfontsize\normalsize{11pt}{13pt}
  \abovedisplayskip      7\p@ \@plus 2\p@ \@minus 5\p@
  \abovedisplayshortskip \z@ \@plus 3\p@
  \belowdisplayskip      \abovedisplayskip
  \belowdisplayshortskip 4\p@ \@plus 3\p@ \@minus 3\p@
}
\normalsize
\renewcommand{\small}{%
  \@setfontsize\small\@ixpt\@xpt
  \abovedisplayskip      6\p@ \@plus 1.5\p@ \@minus 4\p@
  \abovedisplayshortskip \z@  \@plus 2\p@
  \belowdisplayskip      \abovedisplayskip
  \belowdisplayshortskip 3\p@ \@plus 2\p@   \@minus 2\p@
}
\renewcommand{\footnotesize}{\@setfontsize\footnotesize\@ixpt\@xpt}
\renewcommand{\scriptsize}{\@setfontsize\scriptsize\@viipt\@viiipt}
\renewcommand{\tiny}{\@setfontsize\tiny\@vipt\@viipt}
\renewcommand{\large}{\@setfontsize\large\@xiipt{14}}
\renewcommand{\Large}{\@setfontsize\Large\@xivpt{16}}
\renewcommand{\LARGE}{\@setfontsize\LARGE\@xviipt{20}}
\renewcommand{\huge}{\@setfontsize\huge\@xxpt{23}}
\renewcommand{\Huge}{\@setfontsize\Huge\@xxvpt{28}}

\makeatother
\makeatletter
\providecommand{\maketitle}{}
\renewcommand{\maketitle}{%
  \par
  \begingroup
    \renewcommand{\thefootnote}{\fnsymbol{footnote}}
    \renewcommand{\@makefnmark}{\hbox to \z@{$^{\@thefnmark}$\hss}}
    \long\def\@makefntext##1{%
      \parindent 1em\noindent
      \hbox to 1.8em{\hss $\m@th ^{\@thefnmark}$}##1
    }
    \thispagestyle{empty}
    \@maketitle
    \@thanks
  \endgroup
  \let\maketitle\relax
  \let\thanks\relax
}

\newcommand{\@toptitlebar}{
  \hrule height 2\p@
  \vskip 0.25in
  \vskip -\parskip%
}
\newcommand{\@bottomtitlebar}{
  \vskip 0.29in
  \vskip -\parskip
  \hrule height 2\p@
  \vskip 0.09in%
}

\providecommand{\@maketitle}{}
\renewcommand{\@maketitle}{%
  \vbox{%
    \hsize\textwidth
    \linewidth\hsize
    \vskip 0.1in
    \@toptitlebar
    \centering
    {\LARGE\sc \@title\par}
    \@bottomtitlebar
    \textsc{}\\
    \vskip 0.1in
    \def\And{%
      \end{tabular}\hfil\linebreak[0]\hfil%
      \begin{tabular}[t]{c}\bf\rule{\z@}{24\p@}\ignorespaces%
    }
    \def\AND{%
      \end{tabular}\hfil\linebreak[4]\hfil%
      \begin{tabular}[t]{c}\bf\rule{\z@}{24\p@}\ignorespaces%
    }
    \begin{tabular}[t]{c}\bf\rule{\z@}{24\p@}\@author\end{tabular}%
  \vskip 0.4in \@minus 0.1in \center{\today}   \vskip 0.2in
  }
}
\makeatother
\providecommand{\keywords}[1]{\textbf{Keywords: } #1}

\usepackage[verbose=true,letterpaper]{geometry}
\title{Intuition First or Reflection Before Judgment? The Impact of Evaluation Sequence on Consumer Ratings}

\author{
 He Wang \\
  School of Business\\
  Renmin University of China\\
  \texttt{he.wang@ruc.edu.cn} \\
   \And
 Yueheng Wang \\
  School of Business\\
  Renmin University of China\\
  \texttt{yuehengw@163.com} \\
\And
 Ziyu Zhou \\
  School of Statistics\\
  Renmin University of China\\
  \texttt{ziyu.zhou@ruc.edu.cn} \\
\And
 Hanxiang Liu \\
  School of Business\\
  Renmin University of China\\
  \texttt{hanxiang.liu@ruc.edu.cn} \\
}

\begin{document}
\onehalfspacing
\maketitle

\begin{abstract}
As online reviews increasingly drive consumer decisions, the impact of review interface design on rating authenticity remains under-explored. This research investigates how evaluation sequence ("Rating-First" vs. "Review-First") influences consumer ratings through three experiments and a large-scale secondary data analysis. 
The results reveal a significant polarization effect: in high-quality service contexts, the "Rating-First" sequence (vs. "Review-First") increases overall ratings, whereas in low-quality contexts, it leads to significantly lower ratings. This mechanism is driven by a serial mediation path of affective heuristics and cognitive effort. Furthermore, product attributes moderate this effect, with hedonic products amplifying the rating extremity compared to utilitarian ones. Secondary data from Yelp and Letterboxd confirm these findings, showing that the "Rating-First" platform (Yelp) exhibits a polarized bimodal distribution, while the "Review-First" platform (Letterboxd) shows more concentrated ratings.
In conclusion, this research reveals how evaluation sequence shapes consumer ratings through affective and cognitive paths from an information-processing perspective. These findings extend the theoretical understanding of the online review formation process and offer practical insights for platforms to optimize interface design and enhance rating authenticity and credibility.
\end{abstract}

\keywords{Review Sequence and Ratings, Affective Heuristics, Cognitive Effort, Product Attributes, Behavioral Experiments, Secondary Data Analysis}

\newpage

\section{Introduction}
In recent years, the rapid expansion of the digital platform economy has established online consumer review systems as a vital information mechanism connecting consumers, platforms, and merchants. Grounded in the principle of social proof, individuals often reference the behaviors and evaluations of others to reduce decision-making uncertainty when lacking sufficient information \citep{craik1972levels}. As online review systems continue to evolve, the variations in review processes and interaction methods across different platforms have emerged as significant contextual factors shaping consumer evaluation behavior.

Existing research on online reviews primarily adopts a consumer-side perspective, treating reviews and ratings as exogenous information cues and focusing on how they influence consumer judgment and decision-making. A vast body of literature starts from the characteristics of review content to explore the impact of text length, emotional valence, and argumentative structure on review helpfulness and persuasiveness. For instance, \citet{mudambi2010makes} demonstrated that review extremity and depth influence helpfulness, with this impact being moderated by product type; \citet{schlosser2011can} pointed out that the alignment between review argumentative structure and ratings affects persuasiveness. Furthermore, studies have shown a marginal diminishing effect of review length, while social validation cues, such as historical helpfulness records, also influence consumer judgment \citep{huang2015study}. Recently, some researchers have begun to focus on how the presentation order of reviews affects information-processing paths, finding that different sorting methods can alter systematic and heuristic processing, thereby influencing review helpfulness and purchase intention \citep{lee2022exploring}. Overall, this research tradition mainly focuses on how reviews serve as information inputs to influence consumer decision-making.

In contrast, research on the generator-side mechanisms of reviews and ratings remains relatively limited. Literature from the generator-side primarily explores consumers' motivations for posting reviews \citep{hennig2004electronic} and the reasons behind the polarized trends in rating distributions. Studies have pointed out that online ratings are often subject to self-selection bias, as only a subset of users choose to participate in the evaluation process, resulting in rating distributions that exhibit "J-shaped" or extreme characteristics \citep{hu2017self,moe2012online}. However, most studies in this vein explain rating distributions from the perspectives of participation choice and user heterogeneity, paying less attention to how platform interface design influences the psychological processing of consumers during the evaluation generation process.

Therefore, although existing literature has uncovered important mechanisms of online reviews from both the consumer and generator perspectives, a systematic empirical investigation remains absent regarding how the specific interface design factor of "Rating-Review Sequence" influences final ratings and their degree of extremity by altering consumers' affective and cognitive processing paths.
To bridge these gaps, this study focuses on the mechanism through which review order influences overall consumer ratings, employing a multi-method approach that combines consumer behavior experiments with secondary data analysis. Through experimental designs, we examine the impact of review order on overall ratings and construct a model incorporating two mediating variables: degree of affective heuristics and degree of cognitive effort, while introducing product attributes as a moderating variable.
Drawing upon Dual-process Theory \citep{kahneman2011thinking}, individual judgment is simultaneously influenced by a fast, intuitive affective processing system (System 1) and a slow, analytical cognitive processing system (System 2). This study posits that a "Rating-First" interface is more likely to activate rapid processing reliant on intuition and emotional responses, thereby amplifying the role of affective heuristics in evaluation formation. In contrast, a "Review-First" interface may prompt consumers to engage in more thorough information integration and reflective judgment, strengthening the role of cognitive processing in rating formation, which ultimately exerts a systematic impact on the final rating. Finally, this study analyzes approximately 7.7 million consumer review records from Yelp (where users rate before reviewing) and Letterboxd (where users review before rating) to compare the differences in rating distributions under these two sequences, thereby enhancing the external validity of our conclusions in real-world contexts.

\section{Related Work}
\subsection{Consumer Review Sequence}
\citet{arndt1967role} was among the first to note that positive evaluations make consumers more inclined to accept new products, while negative evaluations have the opposite effect. This work systematically discussed the role of consumer interpersonal communication in new product diffusion and consumer decision-making, attributing significant value to consumer evaluations and prompting subsequent research to focus on improving or controlling the objectivity of such ratings. \citet{mcfarland1981effects} pointed out that the order of questions in a questionnaire can alter respondents' reactions to subsequent items. In the context of quantitative surveys, \citet{schwarz1991assimilation} proposed the assimilation and contrast effects, suggesting that when specific questions precede general ones, the responses to the latter are influenced by the activation of the prior context. Similarly, \citet{garbarski2015effects} noted that the sequence of question presentation systematically affects subjects' responses; specifically, answering specific items first strengthens the correlation between those items and the overall evaluation.

Regarding the sequence of consumer reviews, research indicates that consumer satisfaction is shaped by the integration of multiple service cues. When an overall attribute assessment precedes a quantitative satisfaction evaluation, the variance in consumer product satisfaction can be better explained \citep{auh2003order}. Overall, the integrative process of consumer satisfaction is sensitive to sequence; performing an objective evaluation first tends to make the differences in the integration process more pronounced. \citet{eryarsoy2014experimental} pointed out that the biased presentation order of online product reviews can lead consumers to assign excessive weight to early reviews, triggering effects that diverge from their initial intentions and subsequently impacting systematic decision-making. Similarly, \citet{mehr2024does} argued that information presentation and review order significantly influence final rating behavior; specifically, rating sub-attributes before providing a global rating significantly alters a consumer's final overall satisfaction score. However, these studies have not investigated the sequential effects between textual reviews and numerical ratings, leaving a research gap that this study aims to address.

Further research has identified a first-order positive sequential dependence in scale responses, which leads to a decrease in the true reliability and validity of subscales. Notably, under grouped presentation conditions, estimates of reliability and validity are higher than those under random presentation. These findings suggest that subsequent studies should employ explicit modeling of sequential dependence to more accurately assess scale reliability \citep{shimada2023sequential}. Additionally, recent experimental evidence indicates that in "rating-first" scenarios, overall scores tend to be higher compared to scenarios where detailed evaluations precede the rating, directly revealing the discrepancy between scores generated under different sequences \citep{duffek2024review}. However, that study did not control for contexts involving high-quality service levels, leaving significant room for further exploration.

\subsection{Affective Heuristics and Cognitive Effort}
\citet{garbarino1997cognitive} investigated whether the exertion of cognitive effort by consumers triggers negative emotions and alters choice outcomes. Their findings revealed that as cognitive effort increases, consumers' negative affect rises significantly; consequently, when faced with options yielding similar outcomes, consumers tend to favor the alternative that evokes less negative emotion. \citet{westbrook2013subjective} further noted that individuals perceive cognitive effort as a resource-depleting activity, with behavioral assessments being significantly influenced by individual differences and varying mental states. \citet{apps2015role} experimentally confirmed that people exhibit an aversion to tasks requiring high cognitive effort, showing a preference for safe, familiar, and low-risk outcomes. In the context of the present study, the "Review-First" sequence compels consumers to exert greater cognitive effort. Consequently, they may seek to avoid such evaluative modes or, when faced with this requirement, provide lower ratings due to the pessimistic affect generated by perceived cognitive strain \citep{david2024relation}. In summary, the impact of cognitive effort on consumers is predominantly negative, predisposing them toward relatively lower evaluations.

\citet{finucane2000affect} discovered that individuals frequently employ affective heuristics when making risky decisions, assessing risks and benefits based on their immediate emotional responses to stimuli to facilitate choice. Their experiments underscore the significant influence of affective heuristics on consumer decision-making. \citet{king2014affect} demonstrated that under extreme emotional evaluations, consumers' perceptions of risk and products are significantly skewed. Within the framework of this research, when consumers in a high-quality service scenario provide a rating first, their affect tends toward a positive extreme; thus, their scores are elevated by this emotional influence. \citet{ali2016integrated} directly pointed out that superior service attitudes and enhanced emotional experiences lead to higher user satisfaction and a greater willingness to pay a premium. This not only validates the relationship between affective heuristics and user satisfaction but also provides practical guidance for firms to foster customer loyalty and increase profitability. Unlike cognitive effort, affective heuristics typically generate more positive affect, thereby enabling consumers to provide higher ratings.

\subsection{Product Attributes}
As a critical factor influencing consumer behavior \citep{3fa5c73c-7565-38c2-a0d2-e93f88544d84}, product attributes shape the information-processing strategies consumers employ during decision-making and dictate final evaluative outcomes. \citet{oliver1980cognitive} noted that consumers form expectations of a product prior to actual experience; during the evaluation stage, they compare the actual experience with these expectations. The resulting discrepancy influences satisfaction levels and, subsequently, the overall evaluation. Consequently, product attributes themselves frame the consumer's expectancy structure, thereby indirectly influencing subsequent judgmental processes.

Further research suggests that quality assessments are primarily based on cognitive judgments, whereas satisfaction more closely reflects individual emotional reactions during the evaluation process \citep{meirovich2008relationship}. This implies that product attributes not only affect the rational weighing of features but also potentially influence the intensity and direction of emotions experienced during evaluation. Thus, product attributes can alter the relative reliance consumers place on affective versus cognitive information, exerting a systematic impact on the rating formation process.

\citet{overby2006effects} pointed out that in online shopping environments, utilitarian and hedonic values differentially impact consumer preferences and behavioral intentions, emphasizing that the nature of product attributes influences evaluative tendencies and satisfaction. \citet{ryu2010relationships}, through empirical research in the leisure industry, found that both hedonic and utilitarian products significantly affect customer satisfaction. Furthermore, \citet{sen2025influence} provided empirical evidence that, compared to utilitarian products, consumers exhibit stronger emotional engagement with hedonic products and are more likely to generate positive evaluative behaviors and attitudinal outputs.

\section{Hypotheses Development}
\subsection{Theoretical Foundation}
Early research proposed that human information processing consists of two modes: automatic processing and controlled processing. Automatic processing is rapid, occurs without conscious awareness, and consumes minimal resources; in contrast, controlled processing is governed by the subject's consciousness and requires significant attention and cognitive resources. Through experimental studies, it was confirmed that automatic processing forms when stimuli and responses maintain consistency across multiple trials; when inconsistency occurs, controlled processing is employed \citep{schneider1977controlled}. This provided the theoretical underpinning for investigating human thought and information-processing mechanisms. Subsequent domestic and international research has expanded upon this foundation, further refining the mechanisms of these two processing modes.

Regarding automatic processing, scholars have noted that it essentially involves the retrieval of domain-specific knowledge bases, suggesting that practice and consistency influence processing effectiveness according to a power law \citep{logan1988toward}. Further research demonstrated that the incremental gains from repetition vary across items and that these gains stem from the inherent associations between experimental items \citep{logan1990repetition}. In terms of controlled processing, \citet{evans1984heuristic} distinguished between heuristic reasoning and analytic/logical reasoning, establishing clear boundaries for controlled processing and systematically discussing the limitations of cognitive analytic systems in complex reasoning \citep{evans1989bias}. In subsequent studies, \citet{schneider2003controlled} developed the quantitative CAP2 model, explaining the mechanisms of both modes through three modules: perceptual representation, working memory and control, and long-term memory, grounded in neurobiological mechanisms. Later, \citet{evans2008dual} noted in a review that prior research aimed to differentiate between automatic, unconscious cognitive processes and deliberate, conscious ones, mapping the core of various theories into two primary information-processing modes. This laid the groundwork for \citet{kahneman2011thinking} to synthesize existing research into the Dual-process Theory.

The Dual-process Theory, synthesized by Daniel Kahneman in \textit{Thinking, Fast and Slow}, constitutes a vital theoretical foundation for understanding evaluative bias. It posits that human cognition is driven by the interaction between the heuristic System 1 and the analytical System 2, with System 1 predominantly in control most of the time. Empirical research has demonstrated the significant impact of Dual-process Theory on consumer perception, choice, and decision-making. \citet{lu2012application} summarized the application of dual-system decision tools—such as intuitive heuristics and intuitive biases—in promotional contexts, noting that the theory effectively explains various cognitive phenomena during consumer promotional decision-making across multiple levels. \citet{zhang2014examining} investigated the impact of online consumer reviews and found that argument quality and perceived quantity of reviews act as cognitive and heuristic factors, respectively, both exerting significant influence on purchase intentions. \citet{zhang2018fairness} incorporated dual-system thinking into research on fairness preferences in gaming, exploring the mechanism behind individual rejection of unfair distributions and proposing the underlying neural mechanisms. \citet{book2018customer} found that by obscuring base-rate information and increasing the effort required for information processing, consumers' analytical thinking could be prompted to take the lead, thereby enhancing objective evaluation capabilities and reducing reliance on heuristic decision-making. Recent studies have even extended this framework to human value perceptions of AI, indicating that System 2 processing significantly increases perceived AI value compared to System 1 \citep{li2025users}.

Among the extensions of Dual-process Theory, the anchoring effect proposed by \citet{tversky1974judgment} suggests that individuals in uncertain situations rely on the first piece of information encountered as a benchmark. This heuristic processing often leads consumers to tailor their review content to support their initial ratings, potentially compromising objectivity and accuracy in an effort to maintain consistency. Furthermore, the Elaboration Likelihood Model (ELM) \citep{petty1986elaboration} focuses on how cognitive motivation and depth of processing determine whether an individual engages in systematic processing. However, some researchers challenge the dual-system framework, arguing that reasoning should be categorized under a single-system framework, where different types of reasoning emerge from feature grading and varying functional roles of consciousness \citep{osman2004evaluation}. Similarly, \citet{grayot2024dual} argued that Dual-process Theory requires further theoretical justification.

Ultimately, Dual-process Theory asserts that judgment involves both a fast affective system and a slow cognitive system \citep{kahneman2011thinking}. Supporting research indicates that emotional responses often precede cognitive evaluation \citep{zajonc1980feeling} and serve as informational cues influencing subsequent judgments \citep{schwarz1983mood}. In the context of online consumer reviews, the review sequence may alter the extent to which consumers rely on affective versus cognitive processing.

\subsection{Review Sequence and Overall Ratings}
Online review systems typically require consumers to provide both a rating and a written review. However, platforms vary in their execution: some require a numerical rating before the written review, while others mandate the review prior to the rating. Although these two arrangements appear to be mere differences in interface presentation, they may influence final evaluative outcomes by altering consumers' information-processing modes.

Existing research indicates that individuals do not evaluate information solely based on content; rather, they are systematically influenced by the sequence in which information is presented \citep{eisenberg1988order}. Furthermore, studies on online word-of-mouth (eWOM) show that consumers rely on both "informative cues" (such as content quality) and "normative cues" (such as ratings and rankings) when processing evaluations. The information-processing paths corresponding to these different cues significantly affect subsequent judgment and adoption behaviors \citep{filieri2015makes}. This finding suggests that online evaluations are not merely direct reflections of pre-existing attitudes but are likely the result of multi-cue processing within a platform context, providing a empirical basis for exploring how "review sequence" shifts processing paths.

According to Dual-process Theory, individual decision-making is generally influenced by two types of processing mechanisms: one is a rapid, automatic process reliant on emotional and intuitive reactions, while the other is a slow, deliberate process dependent on cognitive analysis \citep{evans2008dual}. The former is more easily activated in contexts of insufficient information or limited decision time, whereas the latter typically requires a higher degree of cognitive investment and information integration \citep{evans2013dual}. When an interface requires consumers to provide a rating first, individuals have often not yet fully integrated their consumption experience information. Consequently, they are more likely to rely on immediate emotional reactions or a global impression for judgment, thereby strengthening the role of intuitive processing in rating formation \citep{schwarz1983mood}. In contrast, if the process requires writing a review first, consumers must recall details of the experience, organize their language, and perform information extraction and integration \citep{craik1972levels}. This process prompts them to exert greater cognitive effort, thereby increasing the weight of rational analysis in the rating decision.

Furthermore, intuitive processing is often based on global impressions or emotional experiences, functioning to simplify information processing and reduce decision costs. However, it is also more prone to amplifying the emotional components of an experience, leading to a stronger tendency toward polarized or extreme evaluations. Conversely, cognitive processing emphasizes the weighing of information and comprehensive assessment, which is more likely to yield relatively cautious and neutral judgments.

Based on the aforementioned analysis, this study proposes Hypothesis H1:

\begin{itemize}
\item \textbf{H1:} The review sequence of "Rating-First, Review-Later" (vs. "Review-First, Rating-Later"), when combined with different service contexts, exerts an \textbf{extremity effect} on overall consumer ratings.
\item \textbf{H1a:} In high-quality service contexts, the "Rating-First, Review-Later" sequence (vs. "Review-First, Rating-Later") leads to higher overall consumer ratings.
\item \textbf{H1b:} In low-quality service contexts, the "Rating-First, Review-Later" sequence (vs. "Review-First, Rating-Later") leads to lower overall consumer ratings.
\end{itemize}

\subsection{The Mediating Mechanism of Affect and Cognition}

Existing research suggests that consumer evaluation is a dynamic psychological construction process involving both emotional responses and cognitive processing \citep{pham2007emotion}. Empirical studies in online review contexts further demonstrate that individuals are simultaneously influenced by informative and normative cues, with different processing paths shaping the final judgment \citep{filieri2015makes}. In evaluative settings, individuals often first generate an immediate emotional experience, which not only forms the basis for initial judgment but also serves as an informational cue that directs subsequent cognitive processing \citep{schwarz1983mood}. Factors affecting processing fluency, such as information presentation formats, may alter an individual's information-processing path by inducing a metacognitive experience of "difficulty" \citep{alter2007overcoming}.

Specifically, when the rating precedes the review, individuals must make a quantitative judgment before information is fully integrated. At this stage, lacking the support of deep cognitive processing, individuals are more likely to rely on immediate emotional experiences and intuitive impressions \citep{kahneman2011thinking}. Conversely, when the review precedes the rating, individuals must first perform linguistic representation and meaning construction of the experience. The depth of cognitive processing invested in this stage directly influences subsequent judgments \citep{lockhart1990levels}. Thus, different review sequences may systematically alter the relative weights of emotion and cognition in judgment formation.

Furthermore, emotion not only directly affects the direction of judgment but also regulates the level of cognitive effort invested. According to the Affect-as-Information Theory, individuals frequently use their emotional state as a basis for judgment and decide whether to engage in further processing based on that state \citep{schwarz1991assimilation}. Generally, negative affect fosters detail-oriented, analytical processing, whereas positive affect promotes global impressions and heuristic processing; the impact of emotion depends on whether it occurs during the encoding or judgment stage \citep{bless1992mood}.

It follows that the initial emotional reaction triggered by the review sequence is likely to influence the depth of subsequent cognitive processing. The dual-process framework posits that when an individual first forms an intuitive judgment, that judgment often serves as a reference point for subsequent cognitive activities, unless additional cognitive resources are invested for correction \citep{evans2013dual}. If the evaluation process requires a rating first, the initial emotional response may be rapidly solidified into a numerical expression, exerting an anchoring effect during the subsequent review writing. This leads individuals to be more inclined toward information filtering and interpretation that aligns with their established judgment, thereby reducing the likelihood of cognitive correction. Conversely, when individuals write a review first, they must construct a linguistic representation by recalling, organizing, and integrating various facets of the experience. This process not only extends processing time but also elevates cognitive engagement, allowing emotional reactions to be further analyzed and regulated.

Furthermore, an extensive body of decision-making research demonstrates that the level of cognitive effort is a critical antecedent variable determining the extremity of a judgment. Lower cognitive effort typically leads individuals to rely on heuristic cues, resulting in more extreme and polarized evaluative outcomes \citep{chaiken1980heuristic}. Conversely, higher cognitive effort facilitates the integration of multifaceted information, rendering judgments more robust and neutral \citep{wegener1999elaboration}. Therefore, if the review sequence alters the level of cognitive effort invested by an individual, the final rating results are likely to undergo systematic changes accordingly. In summary, review sequence may not only directly influence the final rating but may also shape the rating through a continuous psychological processing path—first by influencing emotional reactions and subsequently by affecting cognitive effort.

Based on the theoretical derivation above, this study proposes Hypothesis H2:

\begin{itemize}
\item \textbf{H2:} The degree of affective heuristics and the degree of cognitive effort exert a \textbf{serial mediation effect} on the relationship between the review sequence (in combination with service contexts) and overall consumer ratings, with affective heuristics serving as the first mediator and cognitive effort as the second mediator.
\item \textbf{H2a:} In high-quality service contexts, the "Rating-First, Review-Later" sequence (vs. "Review-First, Rating-Later") positively influences affective heuristics; affective heuristics negatively influence cognitive effort; and cognitive effort, in turn, negatively influences overall ratings.
\item \textbf{H2b:} In low-quality service contexts, the "Rating-First, Review-Later" sequence (vs. "Review-First, Rating-Later") positively influences affective heuristics; affective heuristics negatively influence cognitive effort; and cognitive effort, in turn, positively influences overall ratings.
\end{itemize}

\subsection{The Moderating Role of Product Attributes}
Although the sequence of the evaluation interface may influence consumers' judgment modes, existing research suggests that different product types systematically alter an individual's reliance on affective versus cognitive information. Specifically, consumer attitudes typically comprise two relatively independent components: utilitarian and hedonic, with their respective weights varying systematically across different product categories \citep{voss2003measuring}. For utilitarian products, functionality and practicality are the core values, leading individuals to favor attribute-based trade-offs. In contrast, for hedonic products, emotional experiences and pleasure are more diagnostic, making individuals more likely to form judgments based on the cue of "how do I feel about it?" \citep{chitturi2008delight}. Therefore, product attributes themselves constitute a critical contextual cue that alters the relative reliance consumers place on affective versus cognitive information during the evaluation process.

From the perspective of processing fit, when the information-processing mode activated by the decision environment aligns with the psychological processing pattern corresponding to the product type, individuals are more likely to form a coherent judgment experience and stronger evaluative certainty \citep{chernev2004goal}. For hedonic products, emotional reactions are inherently important evaluative inputs. When cognitive resources are limited or when thorough information integration has not yet occurred, individuals are more likely to rely on automatically activated affective responses rather than deliberate cognitive trade-offs \citep{shiv1999heart}. Consequently, an interface sequence like "Rating-First," which relies more on immediate impressions, is more likely to amplify the polarization trend of ratings. Conversely, for utilitarian products, consumers rely more on systematic analysis and information integration to form evaluations. Thus, even if the interface sequence prompts an earlier rating, that rating is more likely to be grounded in relatively stable attribute trade-offs, thereby weakening the influence of immediate emotions on the final score.

Furthermore, a mismatch between the processing mode induced by the review sequence and the psychological processing required by product attributes may lead to processing conflict and a sense of judgment uncertainty. This, in turn, can reduce evaluative confidence and weaken judgment consistency \citep{aaker2001seek}. Therefore, the intensity of the impact of review sequence on ratings likely depends on the information-processing context defined by the product attributes.

\begin{figure}[htbp]
\centering
  \includegraphics[width=\textwidth]{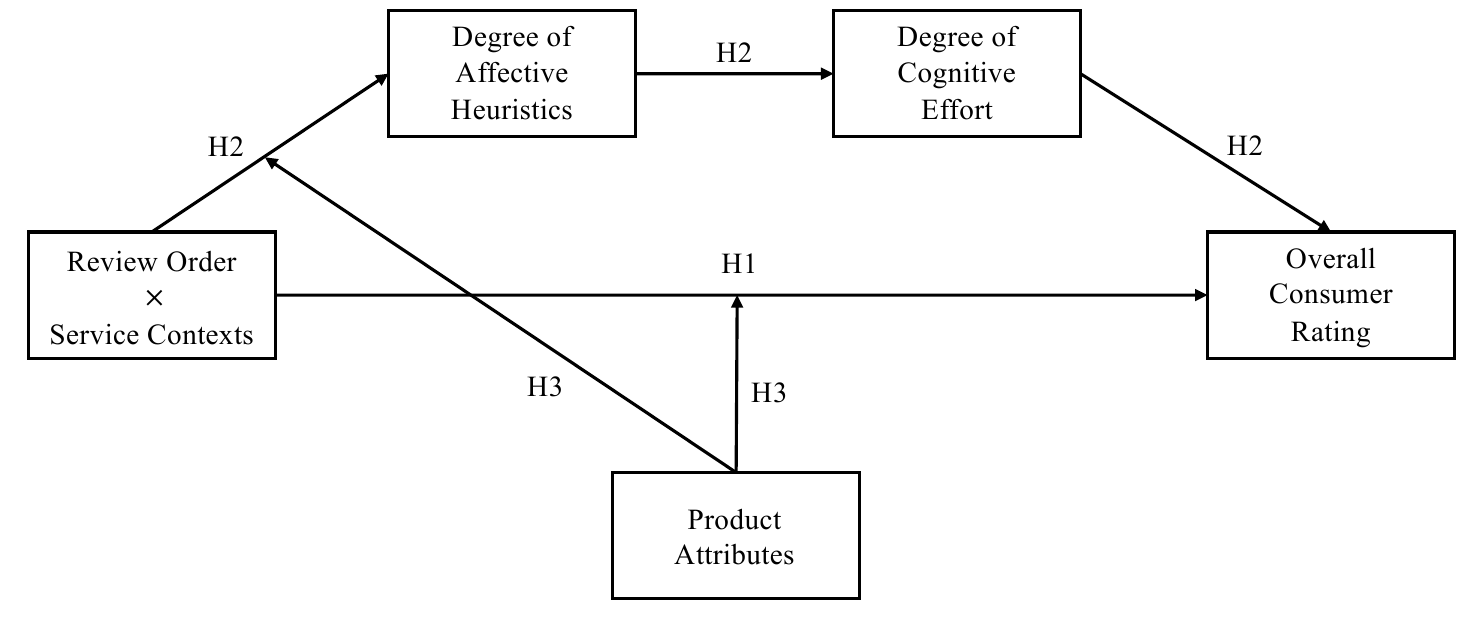}

\caption{Reserch Framework.}
\label{fig:framework}

\end{figure}

In summary, this study proposes Hypothesis H3 and establishes the research framework as shown in Figure \ref{fig:framework}.

\begin{itemize}
\item \textbf{H3:} Product attributes moderate the impact of the review sequence (in combination with service contexts) on overall ratings. Specifically, in hedonic (vs. utilitarian) product contexts, the extremity effect of the "Rating-First, Review-Later" sequence on overall ratings is stronger.
\item \textbf{H3a:} In high-quality service contexts, the positive impact of the "Rating-First, Review-Later" sequence on overall ratings is stronger for hedonic products compared to utilitarian products.
\item \textbf{H3b:} In low-quality service contexts, the negative impact of the "Rating-First, Review-Later" sequence on overall ratings is stronger for hedonic products compared to utilitarian products.
\end{itemize}

\section{Research Design and Results}

\subsection{Experiment 1: The Relationship Between Review Sequence and Overall Consumer Ratings}

\subsubsection{Experimental Design}
The primary objective of Experiment 1 was to verify the impact of the sequence of rating and reviewing on overall consumer ratings across two distinct service contexts (H1). We recruited 240 participants via the Credamo platform. Experiment 1 employed a $2\times 2$ between-subjects design (Review Sequence: Rating-First vs. Review-First $\times$ Service Quality: High-quality vs. Low-quality). Following the methodology of \citet{oppenheimer2009instructional}, we incorporated an attention check item (i.e., "Please do not answer this question"); participants who selected this item were excluded from the analysis. After removing 21 participants who did not complete the questionnaire attentively, a final valid sample of 219 participants was obtained (53\% female, $M_{\mathrm{age}}=28.36$). Among them, 112 were assigned to the Rating-First group (60 in the high-quality context and 52 in the low-quality context), and 107 were assigned to the Review-First group (49 in the high-quality context and 58 in the low-quality context). Drawing on prior research, this experiment manipulated the process using visual comparisons based on textual stimuli. To maximize ecological validity, we utilized a simulated recall method instead of purely hypothetical scenarios.

\subsubsection{Experimental Procedure}
Participants were randomly assigned to one of four groups: Rating-First/High-quality, Review-First/High-quality, Rating-First/Low-quality, or Review-First/Low-quality. All four groups were presented with an experimental interface containing textual stimuli. After confirming they had read and processed the stimuli (with a minimum required time of 10 seconds), they entered a simulated dining evaluation platform.

The specific content of the Textual Stimuli is as follows:

\begin{tcolorbox}[colback=gray!5,colframe=gray!50,arc=10pt,outer arc=10pt,left=15pt,right=15pt,top=10pt,bottom=10pt]
\small
Please take a moment to recall your most recent dining experience—one that left you feeling \textbf{satisfied (or dissatisfied)}. This may have occurred at a restaurant, a café, or any dining establishment. In your mind, please reconstruct the scene as vividly as possible:

\begin{itemize}
\item \textbf{Recall:} What was the environment and atmosphere of that restaurant like?
\item \textbf{Remember:} Think about the items or dishes you ordered. How were their appearance, aroma, and taste?
\item \textbf{Reflect:} How was the quality of service provided by the staff?
\end{itemize}

Please ensure this represents a complete experience in your memory. In the next section, we will simulate a real dining evaluation platform. Please complete the evaluation tasks based specifically on this particular experience.
\end{tcolorbox}

Subsequently, participants provided an overall evaluation of the service based on the details they recalled. The interface design was used to strictly enforce the sequence: either Rating-First or Review-First. During the rating task, participants evaluated the service on a 5-point Likert scale (1=Very Dissatisfied, 5=Very Satisfied). During the reviewing task, participants were required to write an objective review of at least 30 words. Finally, participants completed the attention check and provided demographic information before the experiment concluded.

\subsubsection{Results and Analysis}
\paragraph{Main Effect Test}
Consistent with the predictions in Hypothesis 1, consumers' overall service ratings were more extreme in the "Rating-First" condition compared to the "Review-First" condition. First, a manipulation check was conducted. Using a score of 3 (the scale midpoint) as the baseline, the results showed that the overall ratings in the high-quality service group were significantly higher than 3, while those in the low-quality service group were significantly lower than 3. This indicates that the manipulation of service quality was successful.

\begin{figure}[htbp]
    \centering
    \begin{subfigure}[b]{0.48\textwidth}
        \centering
        \includegraphics[width=\textwidth]{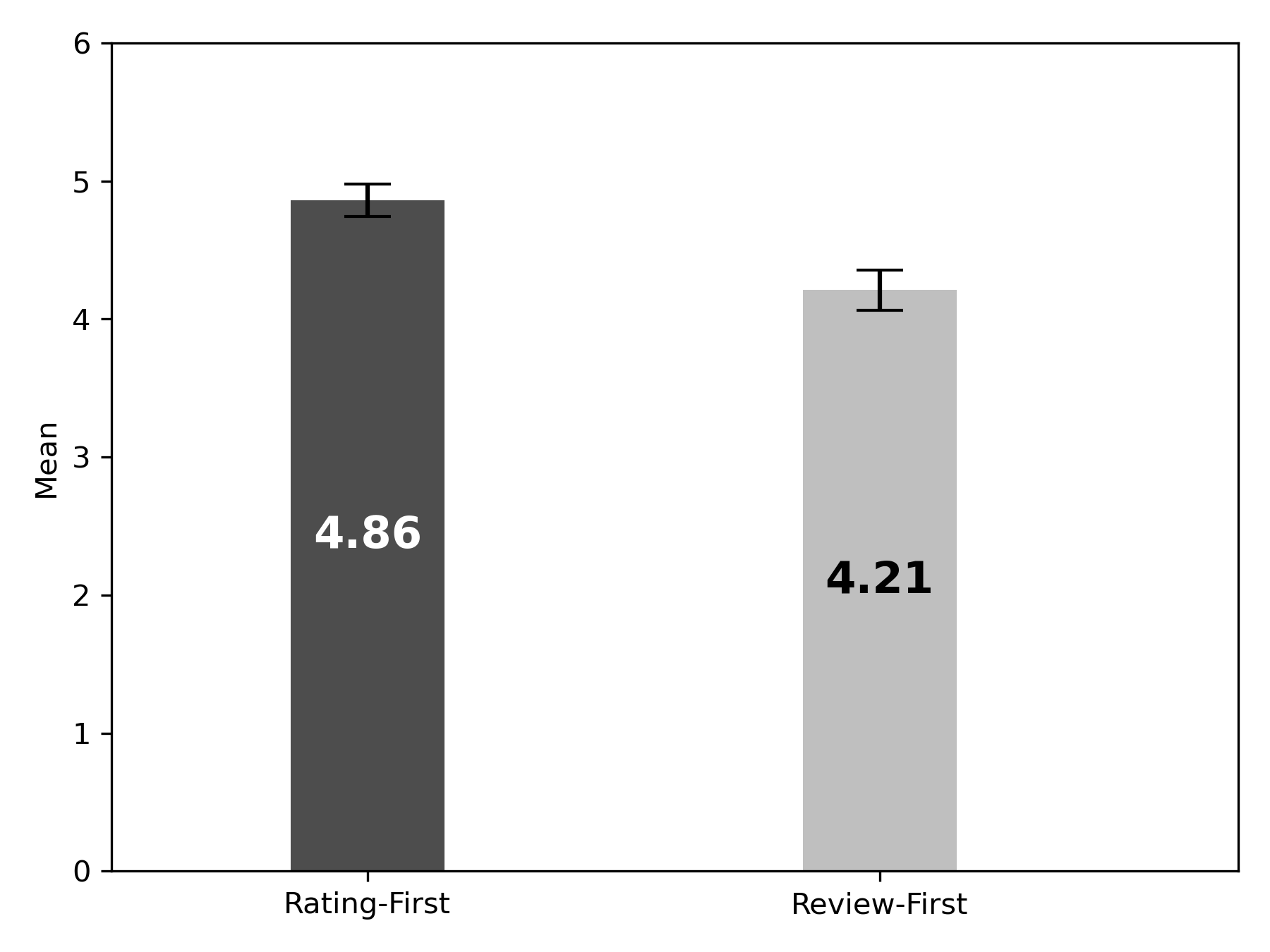} 
        \caption{High-quality Condition}
        \label{fig:good_cond}
    \end{subfigure}
    \hfill 
    \begin{subfigure}[b]{0.48\textwidth}
        \centering
        \includegraphics[width=\textwidth]{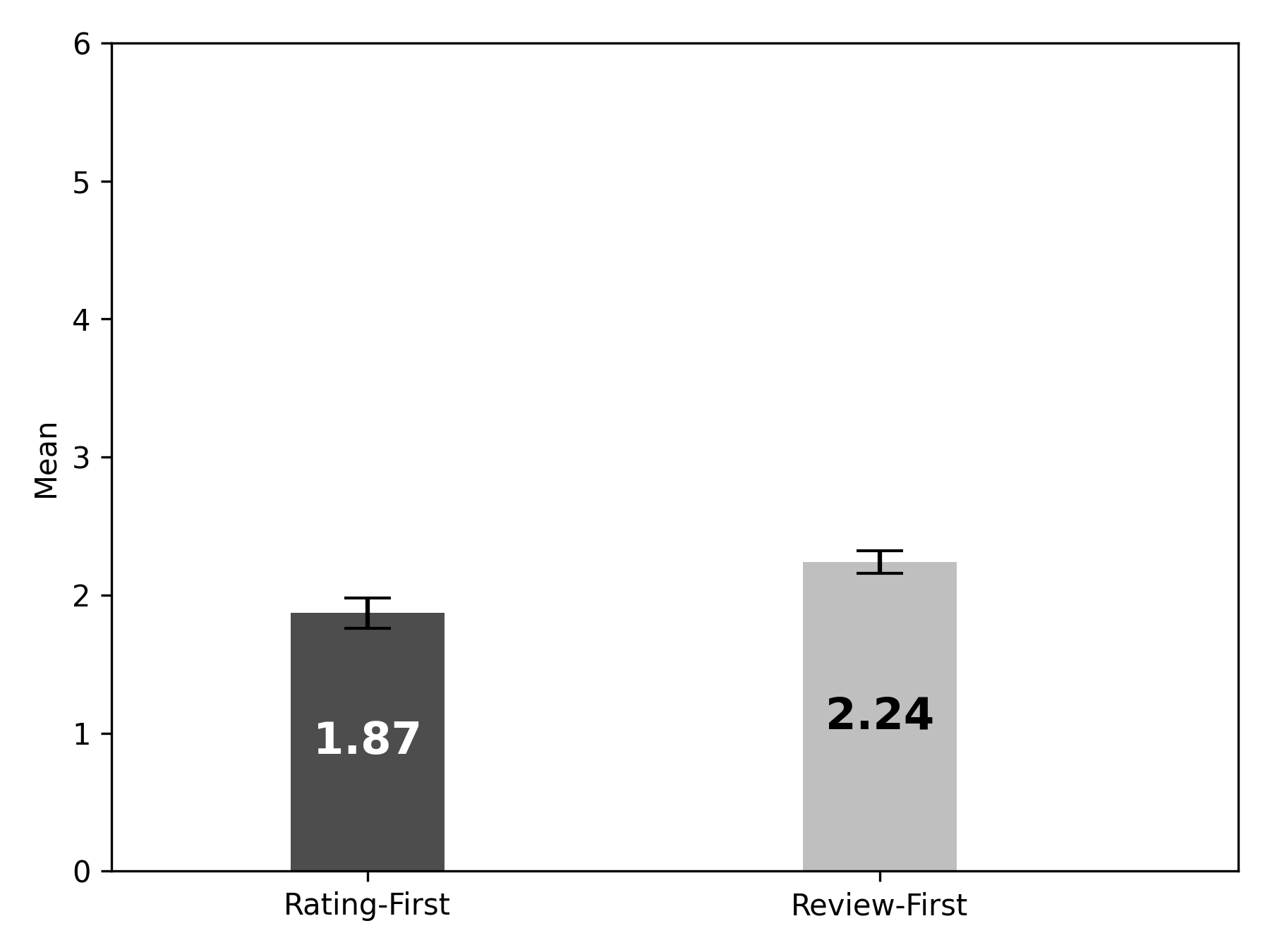} 
        \caption{Low-quality Condition}
        \label{fig:poor_cond}
    \end{subfigure}
    \caption{Error Bar Chart of Dependent Variables under Two Conditions}
    \label{fig:exp_1_error_bars}
\end{figure}

Next, a one-way ANOVA was performed for each of the two service contexts, with the overall rating as the dependent variable. In the high-quality context, participants in the Rating-First group gave significantly higher overall ratings than those in the Review-First group ($M_{\mathrm{Rating-First}}=4.86,SD=0.92 vs. M_{\mathrm{Review-First}}=4.21,SD=1.03;F(1,109)=12.09,p<0.001,\eta^2=0.102$). 
Thus, H1a is supported. In the low-quality context, participants in the Rating-First group gave significantly lower overall ratings than those in the Review-First group ($M_{\mathrm{Rating-First}}=1.87,SD=0.79 vs. M _{\mathrm{Review-First}}=2.24,SD=0.61;F(1,108)=7.64,p<0.01,\eta^2=0.066$). Thus, H1b is supported. Based on these results, Hypothesis 1 is fully supported. The specific results are illustrated in Figure \ref{fig:exp_1_error_bars}.

\paragraph{Summary and Discussion}
The experimental results above indicate that a service interface where ratings precede reviews prompts consumers to give more extreme service ratings. In the subsequent experiment, this study further examines how the sequence of tasks influences consumer ratings through a serial mediation path of cognitive effort and affective heuristics.

\subsection{Experiment 2: The Mediating Role of Cognitive Effort and Affective Heuristics}

\subsubsection{Experimental Design}
The primary objectives of Experiment 2 are twofold: 1) to re-verify the main effect of task sequence (rating vs. reviewing) on consumer ratings; and 2) to test the serial mediation effect of cognitive effort and affective heuristics (H2). Compared with Experiment 1, Experiment 2 utilized front-end rendering for the final evaluation interface to better align with interfaces encountered in daily life, thereby enhancing the study's external validity. 

Experiment 2 employed a 2 $\times$ 2 between-subjects design, identical to Experiment 1. We recruited 260 participants via the Credamo platform. After excluding 28 participants for inattentive responding, a final valid sample of 232 was obtained (58\% female, $M_{\mathrm{age}}$ = 27.93). The "rating-first" group consisted of 119 participants (59 in the high-quality condition, 60 in the low-quality condition), while the "reviewing-first" group consisted of 113 participants (52 in the high-quality condition, 61 in the low-quality condition).

\subsubsection{Experimental Procedure}
Similar to Experiment 1, participants viewed an experimental interface consisting of textual stimuli. After confirming completion (with a minimum viewing time of 10 seconds), they first provided an overall evaluation of the service, which included both rating and reviewing (the sequence of which was manipulated by group assignment). 

Subsequently, participants completed measurements for cognitive effort and affective heuristics. Based on scales developed in prior literature, the affective heuristics scale consisted of 5 items, while cognitive effort was measured using a single-item scale, as detailed in Table 1.

\begin{table}[htbp]
    \centering
    \caption{Measurement Scales for Experiment 2.}
    \label{tab:scales_exp2}
    \begin{tabularx}{\textwidth}{l X l}
        \toprule
        \textbf{Variable} & \textbf{Measurement Items (Operational Questionnaire)} & \textbf{Source} \\
        \midrule
        \multirow{5}{2cm}{Affective Heuristics} 
        & I mainly relied on my feelings to judge the service quality. & \multirow{5}{3cm}{\citet{shiv1999heart}} \\
        & I carefully analyzed the characteristics of each service aspect when making the overall evaluation. (R) & \\
        & I tended to give ratings based on the features that came to mind most easily. & \\
        & I feel that my evaluation is the result of rational weighting. (R/Adapted) & \\
        & I formed this evaluation with very little deliberate thought. & \\
        \midrule
        \multirow{2}{2cm}{Cognitive Effort} 
        & How much mental effort did you invest in completing the evaluation task? & \multirow{2}{3cm}{\citet{paas1992training}} \\
        \bottomrule
    \end{tabularx}
\begin{flushleft}
\footnotesize
\textit{Note:} (R) denotes reverse-coded items. 
\end{flushleft}
\end{table}
Participants rated the items on a 7-point Likert scale (1 = "Strongly Disagree," 7 = "Strongly Agree" / 1 = "Very Little," 7 = "Very Much"). Finally, they completed an attention check and provided demographic information, marking the conclusion of the experiment.

\subsubsection{Results and Analysis}

\paragraph{Reliability and Validity Tests}
Regarding the measurement scale for affective heuristics, we first conducted reliability and validity tests. Based on the collected data, Cronbach's $\alpha$ was 0.79, indicating acceptable internal consistency. Furthermore, the KMO value was 0.842 (exceeding the 0.5 threshold), and Bartlett's test of sphericity was significant ($p < 0.001$). All items aligned with their predicted factors, with factor loadings exceeding 0.6 and a cumulative variance explained of over 80\%, demonstrating high construct validity. For the single-item scale of cognitive effort, prior literature has established its temporal stability and other psychometric indices; therefore, these tests were not repeated here.

\paragraph{Main Effect Test}
First, the analysis revealed a significant difference in overall consumer ratings between the "rating-before-reviewing" and "reviewing-before-rating" groups. 
In the high-quality condition: $M_{\text{rating-first}} = 4.76$, $SD = 0.91$ vs. $M_{\text{review-first}} = 4.31$, $SD = 1.10$; $F(1, 109) = 5.56, p = 0.02, \eta^2 = 0.049$. 
In the low-quality condition: $M_{\text{rating-first}} = 1.69$, $SD = 0.74$ vs. $M_{\text{review-first}} = 2.01$, $SD = 0.68$; $F(1, 119) = 6.14, p = 0.02, \eta^2 = 0.049$. 
These results provide further support for H1.

\paragraph{Mediation Effect Test}
Next, we tested the serial mediation effect for both service conditions. Using a bootstrap approach with 5,000 resamples, we examined the chain: sequence $\rightarrow$ affective heuristics (M1) $\rightarrow$ cognitive effort (M2) $\rightarrow$ overall rating (DV). The results indicated that the 95\% confidence intervals (CIs) for the indirect effects did not contain zero in either condition. Even after controlling for the mediators, the direct effect of evaluation sequence on ratings remained significant. 

\begin{figure}[htbp]
\centering
  \includegraphics[width=\textwidth]{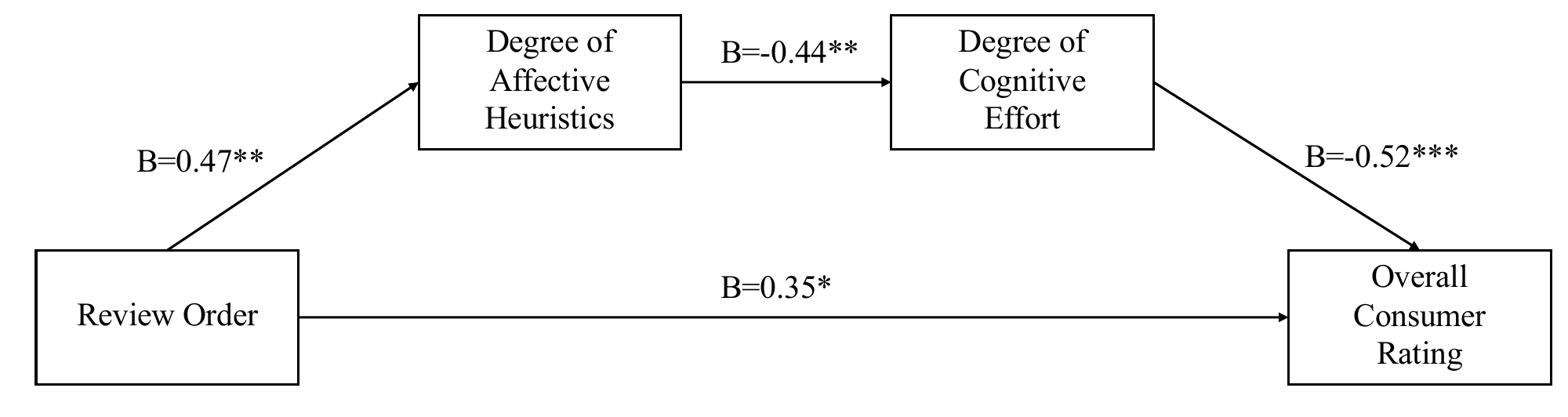}

\caption{The serial mediation mechanism of evaluation sequence on ratings in the high-quality service condition.}
\label{fig:exp2_result_high_quality}

\begin{flushleft}
\footnotesize
\textit{Note:} * $p < 0.05$, ** $p < 0.01$, *** $p < 0.001$. 
\end{flushleft}
\end{figure}

In the \textbf{high-quality condition}, the rating-before-review sequence had a significant positive effect on affective heuristics ($\beta = 0.47, SD = 0.15, t = 3.13, p < 0.01$). Affective heuristics, in turn, had a significant negative effect on cognitive effort ($\beta = -0.44, SD = 0.17, t = -2.59, p = 0.010$). Finally, cognitive effort significantly influenced the overall rating negatively ($\beta = -0.52, SD = 0.15, t = -3.46, p < 0.001$). The indirect effect of this serial path was 0.108 (95\% CI = [0.052, 0.167]). After controlling for these mediators, the direct positive effect of sequence on ratings remained significant ($\beta = 0.35, SD = 0.14, t = 2.48, p = 0.013$), while other indirect paths were non-significant as their 95\% CIs included zero. Thus, H2a was supported (see Figure \ref{fig:exp2_result_high_quality}).

In the \textbf{low-quality service condition}, the rating-before-review sequence had a significant positive effect on affective heuristics ($\beta = 0.42, SD = 0.17, t = 2.47, p = 0.013$). Affective heuristics, in turn, exerted a significant negative effect on cognitive effort ($\beta = -0.29, SD = 0.13, t = -2.23, p = 0.026$). 

\begin{figure}[htbp]
\centering
  \includegraphics[width=\textwidth]{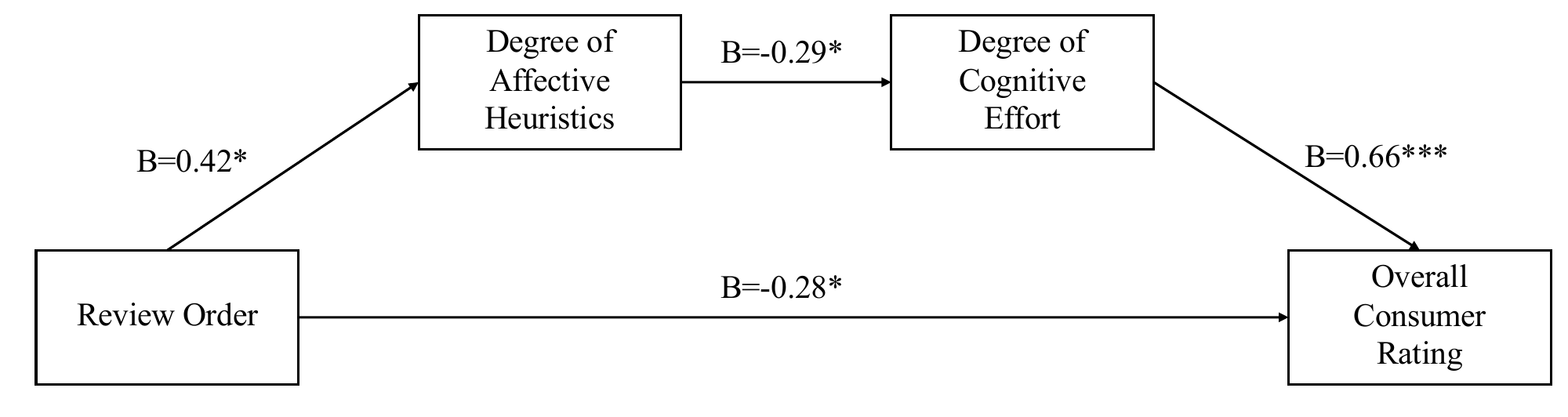}

\caption{The serial mediation mechanism of evaluation sequence on ratings in the low-quality service condition.}
\label{fig:exp2_result_low_quality}
\begin{flushleft}
\footnotesize
\textit{Note:} * $p < 0.05$, ** $p < 0.01$, *** $p < 0.001$. 
\end{flushleft}
\end{figure}

Subsequently, cognitive effort showed a significant positive effect on the overall rating ($\beta = 0.66, SD = 0.20, t = 3.30, p < 0.001$). The indirect effect of this serial mediation path was $-0.080$, with a 95\% CI = [$-0.112$, $-0.047$]. After controlling for these variables, the direct negative effect of evaluation sequence on ratings remained significant ($\beta = -0.28, SD = 0.11, t = -2.55, p = 0.011$). Meanwhile, the 95\% confidence intervals for other indirect paths included zero, indicating they were non-significant. Thus, H2b was supported (see Figure \ref{fig:exp2_result_low_quality}).

These results clarify the specific mechanisms of the serial mediation paths and provide empirical support for H2. Specifically, the findings indicate that a "rating-before-review" sequence consistently heightens affective heuristics regardless of the service context. However, the role of cognitive effort varies: it tends to pull down ratings in high-quality service scenarios, yet "rescues" or buffers ratings to some extent in low-quality ones. This aligns with the "compromise" or "trade-off" logic inherent in this cognitive mode.

\paragraph{Summary and Discussion}
In this experiment, we once again validated the polarizing effect of the "rating-before-review" sequence (compared to "reviewing-before-rating") on consumer scores. Furthermore, we confirmed the partial serial mediating roles of affective heuristics and cognitive effort in this underlying mechanism. In the subsequent experiment, this research will explore the moderating role of \textbf{product attributes} within this influence path.

\subsection{Experiment 3: The Moderating Role of Product Attributes}

\subsubsection{Experimental Design}
The primary objective of Experiment 3 is to explore the moderating role of product attributes on the relationship between evaluation sequence and consumers' overall ratings across different scenarios (H3). According to Hypothesis 3, products with hedonic attributes (vs. utilitarian attributes) strengthen the polarizing effect of the "rating-before-review" sequence (vs. "reviewing-before-rating") on overall scores. 

Experiment 3 employed a $2 \times 2 \times 2$ mixed experimental design, where evaluation sequence and service quality served as between-subjects factors, while product attribute was a within-subjects factor. We recruited 300 participants via the Credamo platform. After excluding 37 participants for inattentive responding, a final valid sample of 263 was obtained (55\% female, $M_{age}$ = 28.64). The "rating-first" group consisted of 137 participants (72 in the high-quality condition, 65 in the low-quality condition), while the "reviewing-first" group consisted of 126 participants (62 in the high-quality condition, 64 in the low-quality condition).

\subsubsection{Experimental Procedure}
The following adjustments were made in Experiment 3:
\begin{enumerate}
    \item All participants experienced both hedonic and utilitarian product conditions sequentially within a controlled environment, with the order of presentation determined by \textbf{counterbalancing} to eliminate order effects. Textual stimuli regarding product category associations were added: 
    \begin{itemize}
        \item \textit{Hedonic condition:} "Recall your most recent dining experience where the primary goal was enjoyment, and you felt satisfied (or dissatisfied)."
        \item \textit{Utilitarian condition:} "Recall your most recent dining experience where the primary goal was to satisfy hunger, and you felt satisfied (or dissatisfied)."
    \end{itemize}
    All other stimuli remained identical to previous experiments.
    \item Compared to Experiments 1 and 2, this study increased the sample size.
    \item After completing each scenario recall, participants performed a perceived product attribute test. Based on prior literature, the measurement of product attributes included 8 items (4 for hedonism and 4 for utilitarianism). These items were presented in a \textbf{cross-randomized} manner to eliminate potential priming or response bias from the Likert scale. Specific items are detailed in Table \ref{tab:product_attributes}.
\end{enumerate}

\begin{table}[htbp]
    \centering
    \caption{Measurement Scales for Product Attributes.}
    \label{tab:product_attributes}
    \begin{tabularx}{\textwidth}{l X l}
        \toprule
        \textbf{Dimension} & \textbf{Measurement Items (Operational Questionnaire)} & \textbf{Source} \\
        \midrule
        \multirow{4}{*}{Hedonism} 
        & This product makes me feel pleasant. & \multirow{8}{3cm}{\citet{batra1991measuring}} \\
        & This product makes me feel happy. & \\
        & I consider this product to be an enjoyment. (Adapted) & \\
        & This product makes me feel good. & \\
        \cmidrule(lr){1-2}
        \multirow{4}{*}{Utilitarianism} 
        & I think this product is useful. & \\
        & I think this product is valuable. (Adapted) & \\
        & I think this product is beneficial. (Adapted) & \\
        & This product is a wise choice. & \\
        \bottomrule
    \end{tabularx}
\end{table}
Participants were asked to rate the items above on a 5-point Likert scale (1 = "Strongly Disagree," 5 = "Strongly Agree").

\subsubsection{Results and Analysis}

\paragraph{Reliability and Validity Tests}
For the affective heuristics scale, Cronbach's $\alpha$ was 0.85, indicating high internal consistency. The KMO value was 0.856 ($> 0.5$), and Bartlett’s test of sphericity was significant ($p < 0.001$). All items aligned with their predicted factors, with factor loadings exceeding 0.6 and a cumulative variance explained of over 80\%, demonstrating strong construct validity.

For the product attribute scale, Cronbach's $\alpha$ was 0.78, confirming its reliability. The KMO value was 0.719 ($> 0.5$), and Bartlett’s test of sphericity was significant ($p < 0.001$). Factor loadings for all items exceeded 0.6, and the cumulative variance explained was over 75\%, indicating that the scale's validity was well-established.

\paragraph{Manipulation Check}
A manipulation check was conducted to compare participants' overall scores on the two attribute dimensions across the hedonic and utilitarian recall scenarios. 
In the hedonic product recall condition, participants scored significantly higher on the hedonism dimension ($M_{\text{hedonic}} = 4.28, SD = 0.76$) than on the utilitarian dimension ($M_{\text{utilitarian}} = 2.85, SD = 0.81$). 
Conversely, in the utilitarian product recall condition, scores were significantly higher on the utilitarian dimension ($M_{\text{utilitarian}} = 3.88, SD = 0.94$) than on the hedonism dimension ($M_{\text{hedonic}} = 1.59, SD = 1.25$). 
These results confirm that the manipulation of product attributes was successful.

\paragraph{Moderation Effect Test}
To test the moderation effect, we employed hierarchical regression analysis, sequentially adding the interaction terms between the moderator and independent variables to the baseline model. We also used the Bootstrap method with 5,000 resamples to calculate the confidence intervals for conditional indirect effects. The results of the moderation effect test are presented in Table 3.
\begin{table}[htbp]
    \centering
    \caption{Moderation Effect Test Results under High-Quality Service Condition.}
    \label{tab:moderation_results}
    \begin{tabularx}{\textwidth}{l c c c l}
        \toprule
        \textbf{Interaction Term} & $\Delta\beta$ & \textbf{SE} & $p$ & \textbf{Significance} \\
        \midrule
        IV $\times$ MO $\rightarrow$ Affective Heuristics & 0.18 & 0.05 & $< 0.001$ & Significant \\
        AH $\times$ MO $\rightarrow$ Cognitive Effort    & 0.12 & 0.09 & 0.186    & Non-significant \\
        CE $\times$ MO $\rightarrow$ DV                  & $-0.09$ & 0.08 & 0.261    & Non-significant \\
        IV $\times$ MO $\rightarrow$ DV                  & 0.13 & 0.06 & 0.015    & Significant \\
        \bottomrule
    \end{tabularx}
\begin{flushleft}
\footnotesize
\textit{Note:} IV = Independent Variable (Evaluation Sequence); MO = Moderator (Product Attributes); AH = Affective Heuristics; CE = Cognitive Effort; DV = Dependent Variable (Overall Rating).
\end{flushleft}
\end{table}

\subsubsection{Conditional Effect Analysis}

\begin{figure}[htbp]
\centering
  \includegraphics[width=\textwidth]{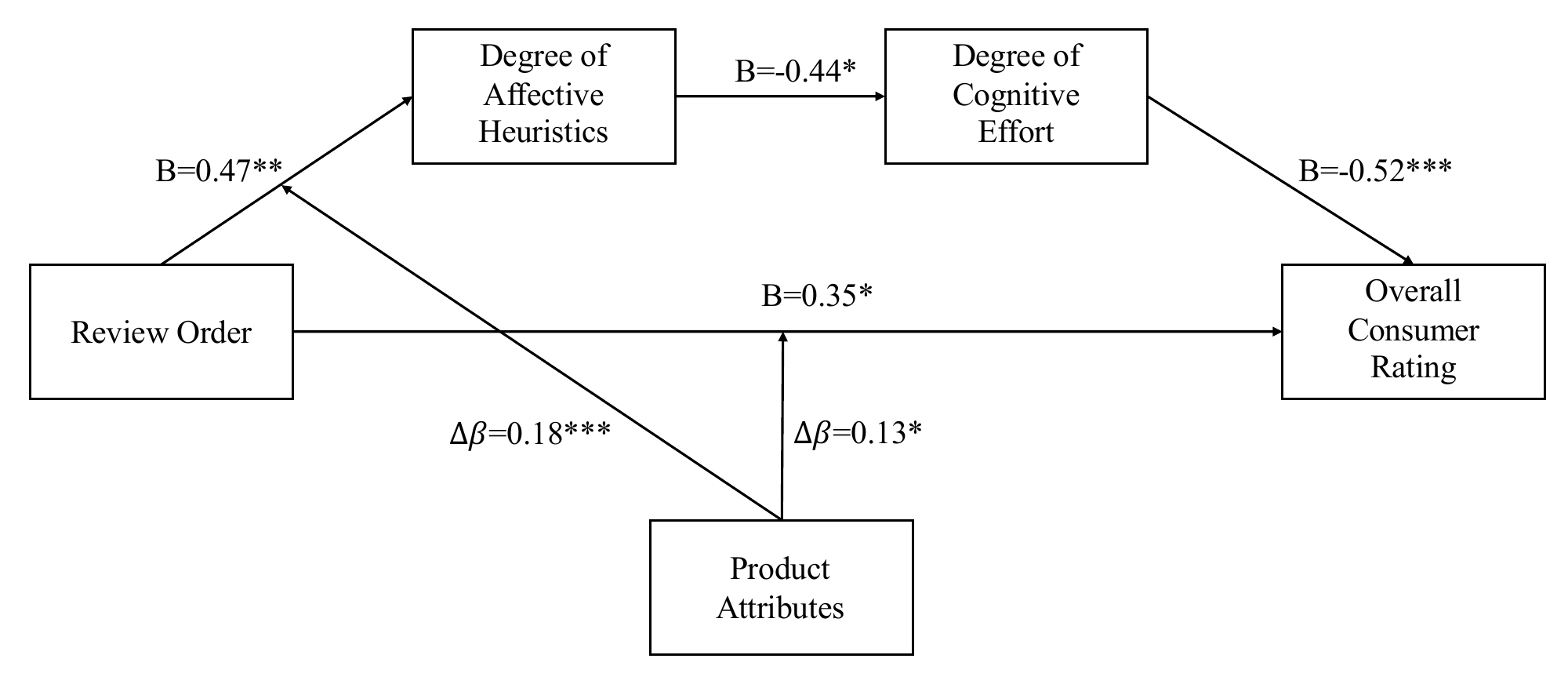}

\caption{The moderating mechanism of product attributes under the high-quality service condition.}
\label{fig:exp3_result_high_quality}
\begin{flushleft}
\footnotesize
\textit{Note:} * $p < 0.05$, ** $p < 0.01$, *** $p < 0.001$. 
\end{flushleft}
\end{figure}

In the high-quality service condition, the test results for two interaction terms were significant, representing the "IV $\times$ MO $\rightarrow$ Affective Heuristics" effect and the "IV $\times$ MO $\rightarrow$ Dependent Variable" effect, respectively. We calculated the differences in conditional effects for the significant paths and used the Bootstrap method to compute confidence intervals. 
The results indicated that when the moderator was at a high level (hedonic products), the positive direct effect of the independent variable on the dependent variable was stronger ($\Delta\beta = 0.13, SE = 0.06, 95\% \text{ CI} = [0.07, 0.19]$). Simultaneously, the positive indirect effect of the independent variable on affective heuristics was also stronger ($\Delta\beta = 0.18, SE = 0.05, 95\% \text{ CI} = [0.14, 0.23]$). Based on these findings, H3a was supported (see Figure \ref{fig:exp3_result_high_quality}).

\begin{figure}[htbp]
\centering
  \includegraphics[width=\textwidth]{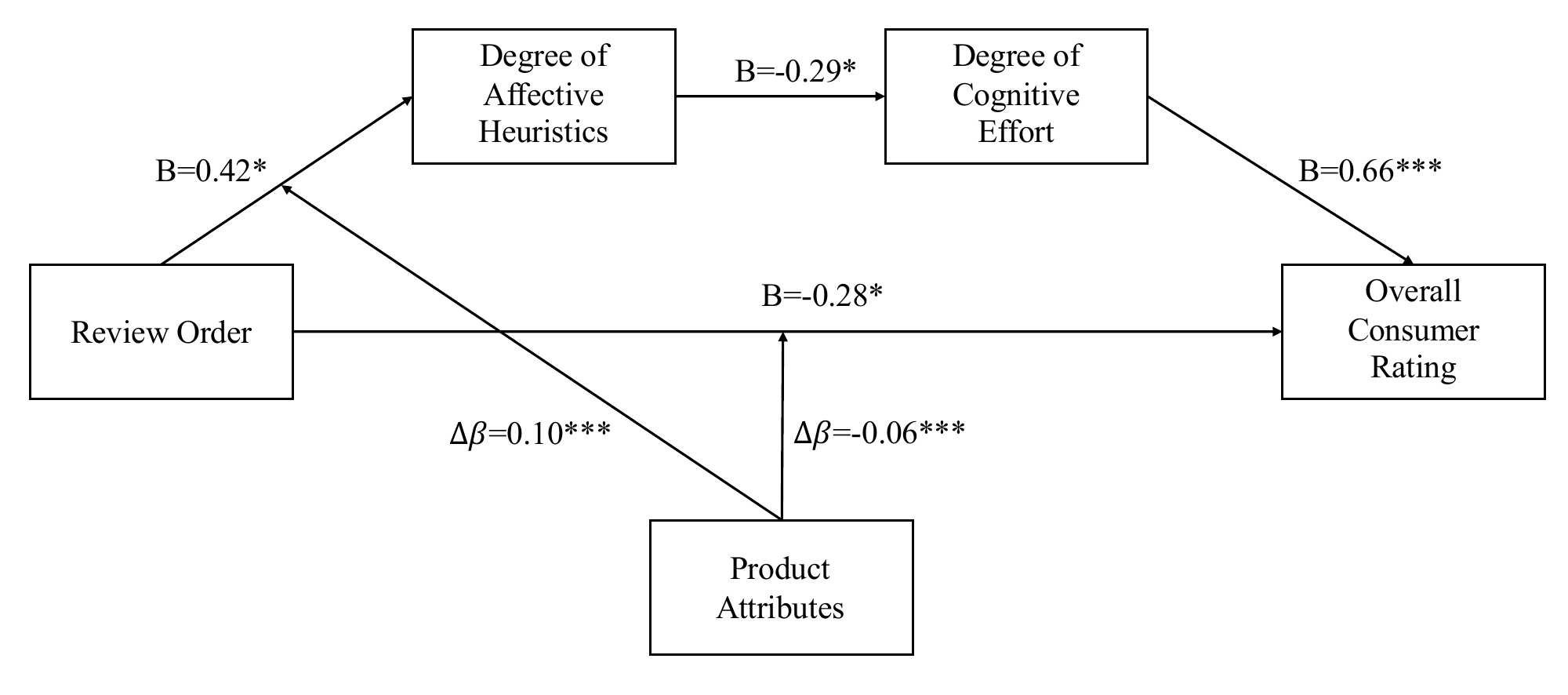}

\caption{The moderating mechanism of product attributes under the low-quality service condition.}
\label{fig:exp3_result_low_quality}
\begin{flushleft}
\footnotesize
\textit{Note:} * $p < 0.05$, ** $p < 0.01$, *** $p < 0.001$. 
\end{flushleft}
\end{figure}

In the low-quality service condition, the significance of the interaction terms remained consistent. We performed conditional effect difference calculations for the significant paths. The results showed that when the moderator was at a high level (hedonic products), the negative direct effect of the independent variable on the dependent variable was stronger ($\Delta\beta = -0.06, SE = 0.01, 95\% \text{ CI} = [-0.07, -0.05]$). At the same time, the positive indirect effect of the independent variable on affective heuristics was stronger ($\Delta\beta = 0.10, SE = 0.03, 95\% \text{ CI} = [0.08, 0.13]$). Consequently, H3b was supported (see Figure \ref{fig:exp3_result_low_quality}).
\paragraph{Summary and Discussion}
In this experiment, we validated the moderating role of product attributes within the underlying influence mechanism. Specifically, when the product is hedonic in nature, the polarizing effect of the "rating-before-reviewing" sequence (relative to "reviewing-before-rating") is significantly amplified. That is, the absolute value of the effect on the dependent variable is strengthened across different service quality scenarios. Based on these findings, Hypothesis H3 was fully supported.

\subsection{Empirical Test Using Secondary Data}

While the aforementioned experiments strictly verified the causal relationships between variables, laboratory environments may be limited by lower external validity. To examine the applicability of these findings in the real world, this study further utilized approximately 7.7 million consumer data points from two platforms, \textbf{Yelp} and \textbf{Letterboxd}, to enhance the ecological validity and generalizability of the conclusions.

\begin{itemize}
    \item \textbf{Yelp} is a platform integrating local business reviews, social sharing, and search. Its core function is discovering and evaluating local merchants. Notably, Yelp follows a \textbf{"rating-before-reviewing"} interface design.
    \item \textbf{Letterboxd}, conversely, is a social networking and review platform centered on movies, facilitating user exchange and sharing of cinematic perspectives. It serves as a rare and typical representative of a \textbf{"reviewing-before-rating"} platform.
\end{itemize}

\begin{figure}[t]
    \centering
    \begin{subfigure}[b]{0.36\textwidth}
        \centering
        \includegraphics[width=\textwidth]{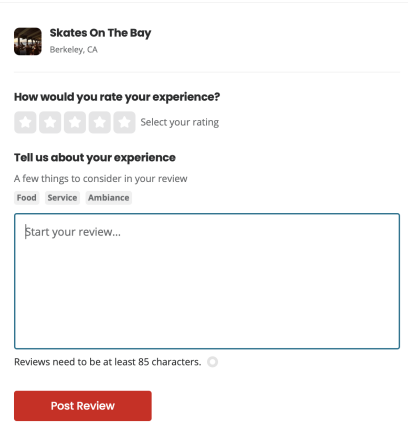} 
        \caption{Yelp}
        \label{fig:yelp}
    \end{subfigure}
    \hfill 
    \begin{subfigure}[b]{0.62\textwidth}
        \centering
        \includegraphics[width=\textwidth]{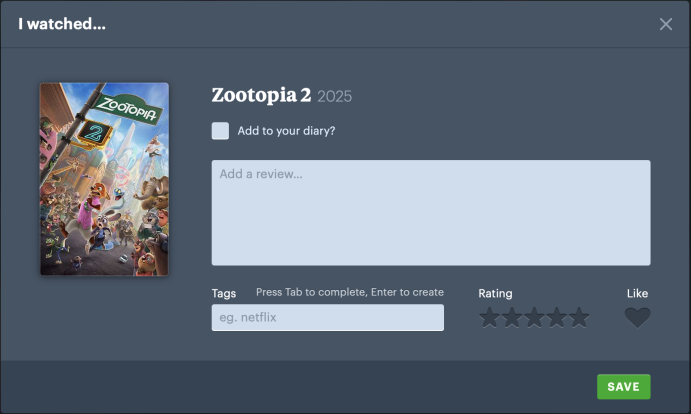} 
        \caption{Letterboxd}
        \label{fig:letterboxd}
    \end{subfigure}
    \caption{Schematic Illustrations of the Evaluation Interfaces}
    \label{fig:score_interfaces}
\end{figure}

In this study, data from Letterboxd were obtained via web scraping, comprising 809,203 rating records for 907 movies. We employed scraping techniques to ensure the \textbf{unbiasedness} of the data. The Yelp data were sourced from its official open dataset, consisting of 6,990,280 rating records. The platform interfaces are illustrated in Figure \ref{fig:score_interfaces}.

\subsubsection{Data Distribution}

\begin{figure}[htbp]
    \centering
    \begin{subfigure}[b]{0.49\textwidth}
        \centering
        \includegraphics[width=\textwidth]{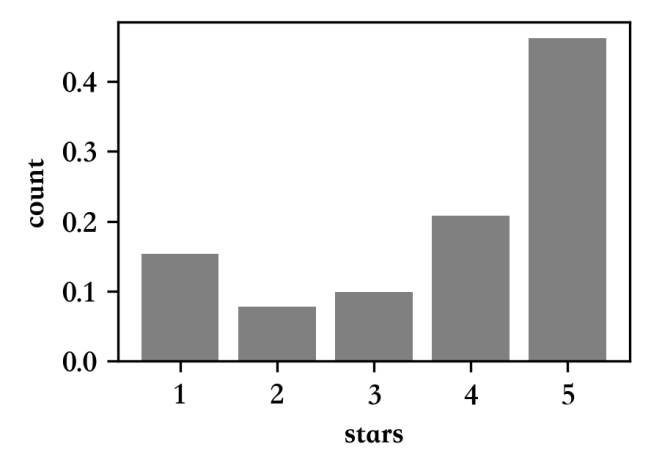} 
        \caption{Yelp}
        \label{fig:yelp_score}
    \end{subfigure}
    \hfill 
    \begin{subfigure}[b]{0.49\textwidth}
        \centering
        \includegraphics[width=\textwidth]{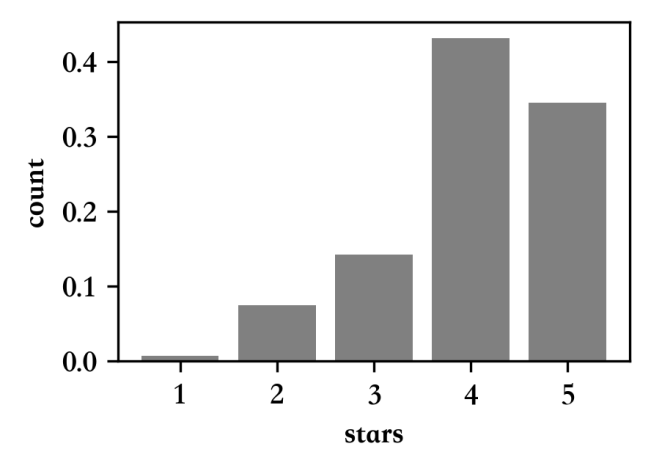} 
        \caption{Letterboxd}
        \label{fig:letterboxd_score}
    \end{subfigure}
    \caption{Score Distribution.}
    \label{fig:score_distribution}
\end{figure}

Based on the analysis of the existing datasets, the results reveal distinct patterns in the rating distributions of the two platforms. 
Yelp ratings exhibit a pronounced U-shaped bimodal distribution. A significant proportion of ratings are concentrated at the extremes—5 stars (high) and 1 star (low)—while the intermediate ratings are relatively sparse. This indicates that user evaluations on Yelp are more polarized.
Letterboxd ratings, conversely, display a right-skewed inverted U-shaped distribution. While users tend to provide relatively high scores, the proportion of extremely high ratings is smaller compared to the majority of neutral or moderately high evaluations.
The distribution of rating data is illustrated in Figure \ref{fig:score_distribution}.

\subsubsection{Statistical Test Analysis}

To further verify the significant differences in distribution characteristics between the two rating mechanisms, we conducted a series of statistical tests on the ratings from Yelp and Letterboxd.
First, the Kolmogorov-Smirnov (KS) test was employed to compare the distributional differences. The results yielded a KS statistic of 0.150 ($p < 0.001$), indicating that the rating data from Yelp and Letterboxd do not follow the same population distribution, both in terms of central tendency and tail shape. 
Additionally, to confirm the differences in proportions across various score intervals, a Chi-square ($\chi^2$) test was performed. The chi-square statistic was approximately 304,715 ($p < 0.001$), further demonstrating significant structural differences in the distribution of the two rating mechanisms.

Finally, to focus specifically on the differences in extreme tail values, we utilized the Anderson-Darling (AD) k-sample test. Compared to the KS test, the AD test assigns higher weights to the tails of the distribution and is more sensitive to tail observations. The calculated Anderson-Darling statistic was 69,757.85 ($p < 0.001$). 

Taken together, the results of all three tests consistently support the conclusion that the "rating-first" and "reviewing-first" mechanisms lead to significantly different rating distributions, particularly regarding extreme values. These findings confirm that the "rating-first" mechanism (Yelp) is more prone to polarized distributions, while the "reviewing-first" mechanism (Letterboxd) tends to produce more concentrated ratings, further validating the robustness of our conclusions.
\section{Conclusion}

Through three controlled experiments and an extensive analysis of secondary data, this research systematically examined the influence mechanism of evaluation sequence on consumer ratings. The findings are summarized as follows:
First, in high-quality service scenarios, a "rating-before-reviewing" sequence leads to significantly higher overall ratings compared to a "reviewing-before-rating" sequence. Conversely, in low-quality service scenarios, the "rating-before-reviewing" sequence results in significantly lower ratings. Overall, this evaluation flow manifests a clear polarization trend.
Second, the study identifies a serial psychological mediation path through which evaluation sequence influences ratings, consisting of affective heuristics and cognitive effort. 
Finally, product attributes play a crucial moderating role in this mechanism. Specifically, compared to utilitarian products, the polarizing effect of the "rating-before-reviewing" mechanism (relative to "reviewing-before-rating") is significantly amplified when the stimulus is a product with strong hedonic attributes.

\subsection{Theoretical Contributions}

At the theoretical level, this research starts from the dynamic process of information processing and reveals the critical role of evaluation sequence—a previously overlooked variable—in the formation of consumer ratings. Since \citet{bass1969new} pioneered the Innovation Diffusion Model, scholars have generally agreed that word-of-mouth (WOM) is a vital bridge connecting consumers and the market. However, past research on WOM and reviews has primarily focused on "what is said" rather than "how it is said." This study addresses this gap by defining the distinction between "rating-first" and "reviewing-first" as a structural behavioral difference, revealing that the evaluation sequence itself acts as a regulatory force in the information processing path.

In the field of online review research, the causes of polarized ratings have long been a focal point of academic discussion. Extensive empirical evidence shows that on platforms such as Amazon, Yelp, and TripAdvisor, rating distributions often exhibit J-shaped or U-shaped characteristics \citep{hu2009online,brandes2022extremity}. Researchers have mostly explained these through the lenses of acquisition bias, under-reporting bias, or differential utility \citep{mudambi2010makes}. While these explanations focus on the composition and motivational differences of consumer groups, they fail to reveal the psychological dynamics of individuals during the act of evaluation. The findings of this study provide a new perspective on this long-standing distributional phenomenon: the polarization of ratings does not stem entirely from external biases but may also be a result triggered by the evaluation process itself. This discovery not only provides new evidence to explain the skewed distribution of online reviews but also reveals how consumer emotion and rationality interact during information processing.

Furthermore, the study points out that product attributes play a moderating role in this mechanism, making the impact of evaluation sequence conditional. This result reveals the interactive relationship between contextual factors and psychological mechanisms, suggesting that consumer emotions do not influence evaluations in isolation but are embedded within specific service contexts. This finding provides an analytical framework for future research that can integrate emotion, cognition, and context.

\subsection{Managerial Implications}

This research provides practical insights with significant guiding value for platform design and corporate management. For high-quality services or products with hedonic attributes, allowing consumers to complete their ratings first while their emotions are still at a peak can capture their immediate positive experiences. This sequence transforms positive affect into visible numerical feedback, thereby strengthening the brand's word-of-mouth effect and market trust. Conversely, in scenarios characterized by unstable service quality or utilitarian products, guiding consumers to write reviews before providing ratings may be a superior approach. This sequence offers users the opportunity to reconstruct their experience through textual organization and expression, thereby reducing the likelihood of emotionalized ratings and making the evaluations more rational and constructive.

Furthermore, the regulatory mechanism of evaluation sequence can be integrated into platform algorithm designs and recommendation systems. Future intelligent evaluation systems could dynamically adjust the order of "rating-first" and "reviewing-first" based on a user's historical behavior, service category, or emotional tendencies to achieve a more authentic and stable rating distribution. This not only helps improve the explanatory power and predictive accuracy of platform data but also enables enterprises to capture signals of emotional fluctuations and satisfaction changes within more nuanced user feedback. The fine-tuning of the evaluation process is, in essence, a means of emotion management and information optimization, returning technical design to a fundamental understanding of human behavior.

By systematically exploring evaluation sequences, this study demonstrates that consumer rating behavior is a form of social expression that is both emotional and cognitive. Understanding this not only helps explain the distribution patterns of ratings but, more importantly, provides new ideas for how to reshape "evaluation credibility" in an era of emotionalization.

\section{Limitations and Future Research}

Although this study systematically examined the influence of evaluation sequence on consumer ratings and its underlying mechanisms, several limitations remain.

First, this research primarily relied on laboratory experiments and recall-based simulations. Although we enhanced external validity through methods such as front-end rendering, differences may still exist between participants' decisions in a simulated environment and their actual evaluation behavior on specific platform interfaces following real consumption. Future research could employ field experiments in collaboration with real-world platforms to test the conclusions of this study in more natural settings, thereby achieving higher ecological validity.

Second, this study focused predominantly on the catering service context. However, service attributes and consumer involvement levels may vary across different industries, such as hotels, e-commerce products, and digital content. The generalizability of the influence mechanism of evaluation sequence in these diverse contexts warrants further verification.

Finally, while this study confirmed that a "reviewing-before-rating" sequence encourages more rational and concentrated ratings, requiring users to write reviews first may increase the barrier to evaluation, potentially leading to a decline in overall response rates. Future research could further explore how to optimize the interface design of the "reviewing-before-rating" sequence. The goal would be to enhance rating quality while minimizing potential negative impacts on user engagement, thereby providing more actionable solutions for platform design.

\newpage

\bibliographystyle{apalike}
\bibliography{references}

\end{document}